\documentclass[aps,prb,preprint,showpacs]{revtex4-1}
\usepackage{graphicx}

\usepackage{amsmath}
\usepackage{amsfonts}
\usepackage{amssymb}
\usepackage{color}

\usepackage{bm}
\usepackage{sidecap}

\usepackage{epstopdf}
\usepackage{hyperref}

\begin{document}


\title{A metamaterial modulator based on electrically controllable electromagnetically induced transparency}
\author{Yuancheng Fan}
\email{phyfan@nwpu.edu.cn}
\affiliation{Key Laboratory of Space Applied Physics and Chemistry, Ministry of Education and Department of Applied Physics, School of Science, Northwestern Polytechnical University, Xi¡¯an 710129, China}
\author{Tong Qiao}
\affiliation{Key Laboratory of Space Applied Physics and Chemistry, Ministry of Education and Department of Applied Physics, School of Science, Northwestern Polytechnical University, Xi¡¯an 710129, China}
\author{Fuli Zhang}
\affiliation{Key Laboratory of Space Applied Physics and Chemistry, Ministry of Education and Department of Applied Physics, School of Science, Northwestern Polytechnical University, Xi¡¯an 710129, China}
\author{Quanhong Fu}
\affiliation{Key Laboratory of Space Applied Physics and Chemistry, Ministry of Education and Department of Applied Physics, School of Science, Northwestern Polytechnical University, Xi¡¯an 710129, China}
\author{Jiajia Dong}
\affiliation{Key Laboratory of Space Applied Physics and Chemistry, Ministry of Education and Department of Applied Physics, School of Science, Northwestern Polytechnical University, Xi¡¯an 710129, China}
\author{Botao Kong}
\affiliation{Key Laboratory of Space Applied Physics and Chemistry, Ministry of Education and Department of Applied Physics, School of Science, Northwestern Polytechnical University, Xi¡¯an 710129, China}
\date{\today}

\begin{abstract}
Electromagnetically induced transparency (EIT) is a promising technology for the enhancement of light-matter interactions, and recent demonstrations of the quantum EIT realized in artificial micro-structured medium have remarkably reduced the extreme requirement for experimental observation of EIT spectrum. In this paper, we propose to electrically control the EIT spectrum in a metamaterial for an electromagnetic modulator. A diode acting as a tunable resistor is loaded in the gap of two paired wires to inductively tune the magnetic resonance, which induces remarkable modulation on the EIT spectrum through the metamaterial sample. The experimental measurements confirmed that the prediction of electromagnetic modulation in three narrow bands on the EIT spectrum, and a modulation contrast of up to 31 dB was achieved on the transmission through the metamaterial. Our results may facilitate the study on active/dynamical technology in translational metamaterials, which connect extraordinary manipulations on the flow of light in metamaterials, e.g., the exotic EIT, and practical applications in industry.
\end{abstract}

\maketitle

Metamaterial, a kind of artificial microstructure with unattainable properties in natural occurring medium, has gained wideattention in physics and material science [1,2]. The arbitrarily designed optical dielectric-functions of metamaterials have been employed for many novel optical phenomena, such as negative refraction [3], diffraction-unlimited imaging [4-6], invisible cloaking [7], subwavelength cavities [8-10], and perfect electromagnetic absorber [11-16] among others. Metamaterials are different from traditional materials for that their responses to external stimuli can be effectively tuned by modeling the geometries of their constituent and thus the local resonant behaviors. However, the local resonant nature of metamaterials limits its operation in a narrow frequency band. It is highly desirable to achieve frequency-agileor multi-band operating metamaterial to extend the working band of a manufactured metamaterial [17-26]. As part of this development, considerable interest has been focused on the realization of actively controlled metamaterials that exhibit tunable optical responsefor practical applications in functional optical devices.

Electromagnetically induced transparency (EIT),originates from the quantum interference between different transition pathways within atoms or molecules coupled to laser fields, was suggested as a technique for eliminating the effect of a medium on a propagating beam of electromagnetic radiation, which can be used to remove optical self-focusing and defocusing, and to improve transmission through an opaque medium in a narrow window with low absorption and steep dispersion [27]. The steep dispersion of EIT is promising for slow light and the enhanced light-matter interactions. However, the extreme environment requirement for quantum EIT obstructs it from practical applications in our daily life. The spectral characteristics of EIT have been reproduced in several classical structures [28-30]. Especially in recent,artificially structured subwavelength elements (namely meta-molecules) [31-38], were realized for the demonstration of EIT spectra in plasmonicmetamaterials. In plasmonicmetamaterials, the EIT is achieved by properly tailoring the coherent interference between two subwavelength resonators in close proximity.

In this paper, we propose and experimentally demonstrate an electrically controlled metamaterial for multi-frequency electromagnetic modulating [18]. We employed a metamaterial comprising of a single wire coupled to a paired wires, a PIN diode acting as tunable medium is loaded in the gap of the paired wires. The physical mechanism of the metamaterial modulator is to manipulate EIT spectrum of the metamaterial by electrically controlling the coupling between the electric mode of a single wire and the magnetic mode of the paired wires. The electrically controlled electromagnetic coupling is implemented by tuning the ¡°on/off¡± state and the resonant strength of the magnetic mode of the paired wires. It was found both theoretically and experimentally that the proposed metamaterial can serve as a modulator for electromagnetic waves at three discrete band located around characteristic peak/dips of the EIT spectrum. Our results may lead to practical applications based on the electrically controlled electromagnetically induced transparency or other classic analogues to quantum interference phenomena in coherent media [39].

The schematic of the proposed EIT metamaterial is shown in Fig. 1(a), the metallic structure is positioned on a 1 mm-thick dielectric substrate with relative electric permittivity of 2.65. A 35 $\mu$m-thick copper film is patterned to metamaterial design with EIT-like spectral response, the structure is composed of a single warped wire (left) coupled with two paired warped wires (right). The width and height of the substrate are w=22.14 mm and $\ell$=47.54 mm, and the width and height of the left wire are $w_{1}$=16 mm and ${\ell_1}$=5mm, the width and height of the right wires are $w_{2}$=12 mm and ${\ell_2}$=17 mm, the metal wires are all with line-width of $w_{\ell}$=1mm, a gap of g=1mm between the paired wires is designed for the loading of a PIN-diode (model: SMP1345). We performed full-wave numerical simulations with a Finite-Difference-Time-Domain (FDTD) based electromagnetic solver for the electromagnetic response of the EIT metamaterials.

\begin{figure}[h]
\centering\includegraphics[width=3.0in]{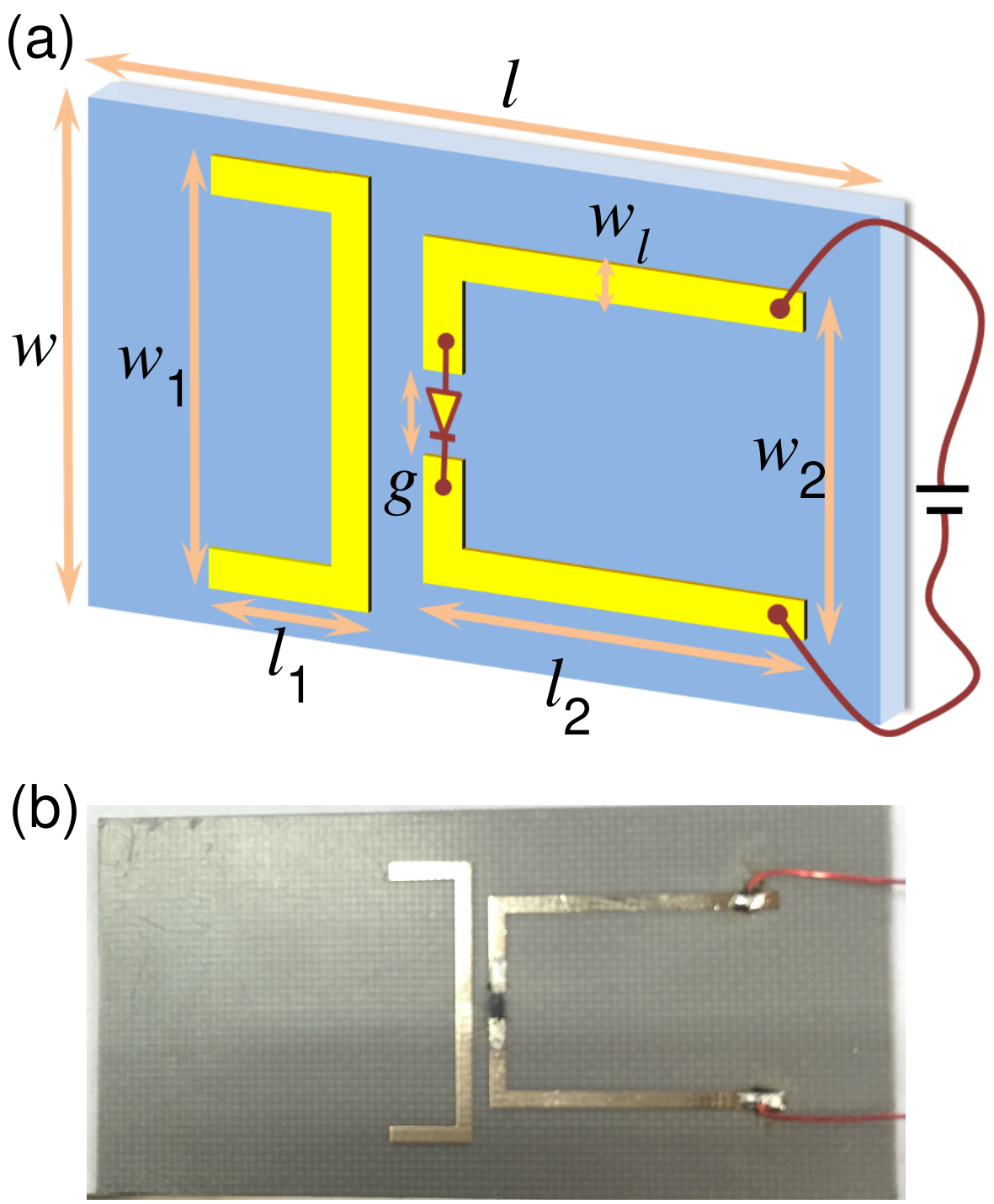}
\caption{(a) Schematic illustration of the proposed metamaterial switch composed of a single wire ( left) coupled with a wire pair (right), a PIN diode is located in the gap of the wire pair for active modulation, geometric parameters of the metamaterial are denoted with black letters. (b) Photograph of a sample fabricated for experimental measurement.}
\end{figure}

We first consider the spectral response of the coupled metamaterial design, the calculations were carried out within a PEC surrounded air box (to simulate our experimental setup: a C-band waveguide). For a metamaterial with only the left wire, it is shown that there is a resonant dip around 5.1 GHz on the transmission spectrum [see in Fig. 2(a)], the resonance is a broad electric-dipolar mode from our investigation on the local-fields. The dipolar mode is similar to the bright mode of the wire in the plasmonic induced transparency (PIT) study [31], which provide an opaque spectral range for the formation of EIT. Then we consider the property of the paired wires, the transmission spectrum is presented by the green curve in Fig. 2(a), in which a sharp resonance around 5.0 GHz is observed, the sharp resonance was verified to be a magnetic-dipolar mode from the local-field investigations (not presented here). Finally, we simulated the case of a metamaterial comprised of both the left single wire and the right wire pair, the transmission spectrum is presented in Fig. 2(a), it can be seen that the original transmission dip of a wire is now changed to a transmission peak at the same frequency with the introduction of the wire pair, and the transparency is accompanied by two transmission dips on either side of the peak, these peak/dips are just the physical picture of constructive and destructive interferences in EIT. The induced surface current distribution at the peak frequency is shown in Fig. 2(b), in which we can see that the electric-dipolar mode of the single wire and magnetic-dipolar mode of the wire pair were both well excited and together contributed in the transmission peak within the opaque band of original electric mode.

\begin{figure}[h]
\centering\includegraphics[width=4.71in]{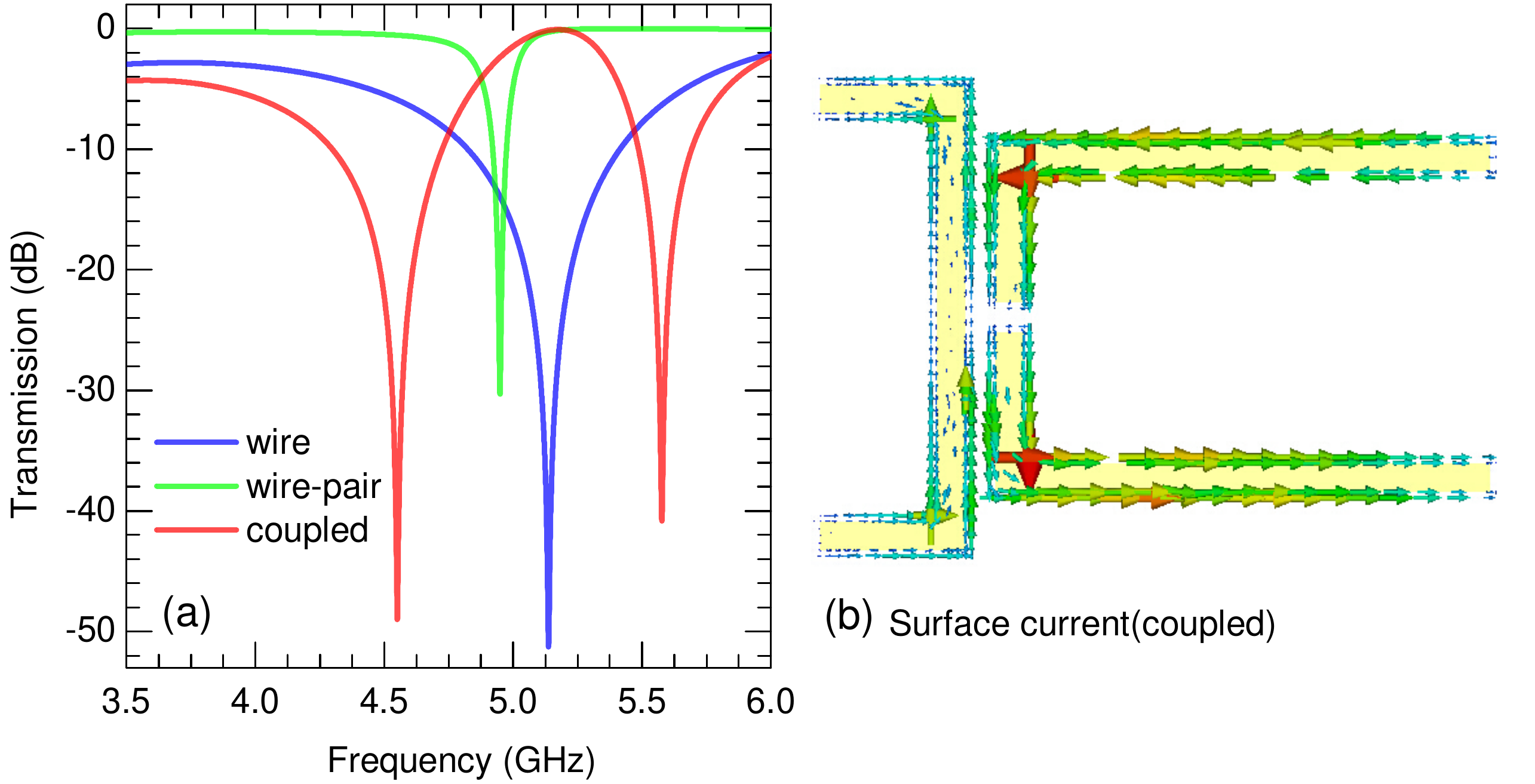}
\caption{(a) Calculated transmission (in dB) spectra of a single warped wire (blue solid), a pair of warped wires (green solid), and the metamaterial with coupled single wire and wire pair (red solid). (b) Surface current distribution of the metamaterial with coupled single wire and wire pair at the peak frequency of 5.17GHz.}
\end{figure}

It is known that the EIT spectrum results from the coherent interference between a sharp resonant and relatively broad resonant background mode, and the sharp resonance is crucial in the formation process of EIT which should be designed: i) with sharp phase changing compared to the electric resonant background for simultaneously achieving destructive and constructive interferences in a narrow frequency range; ii) of proper spectral-overlap with respect to the resonance of electric mode of the single wire. In this study, we propose to realize an electrically biased metamaterial modulator at multi-frequencies by taking advantages from the second recipe of formation EIT and the multi-peak/dip spectrum of EIT. Specifically, we suggest to electrically tuning the conductivity in the gap of paired wires (see Fig. 1)  to inductively turn the resonance [see Fig. 2(a)] within the resonant regime of the left single wire. Our numerical results indicate that the magnetic resonance of the wire pair can be shift to higher frequency (out of the electric resonance of the left wire) by connecting the two discrete wires metallically (results are not shown here), the formation of EIT can thus be interdicted consequently for realizing electromagnetic modulator at specific frequencies.

Figure 3(a) present the simulated transmission spectra (in dB) of different inductance between the paired wires, that is a serious of lumped resistance in the gap, the resistance was changed from 7000$\Omega$(unconnected) to 10$\Omega$(connected).  As shown in Fig. 3(a), the initial EIT spectrum for high resistance gradually changes to resonant spectrum of the electric wire the resistance decreasing to small values like 70$\Omega$, 40$\Omega$, and 10$\Omega$. The EIT peak around 5.1 GHz vanished and varied to a transmission dip of about -35 dB. The two resonant dips associated with the EIT peak around 4.6 and 5.6 GHz also vanished and varied to be of high transmission. These significant modifications on the spectra at three discrete narrow bands are promising for high efficiency electromagnetic modulating application.

\begin{figure}[h]
\centering\includegraphics[width=3.5in]{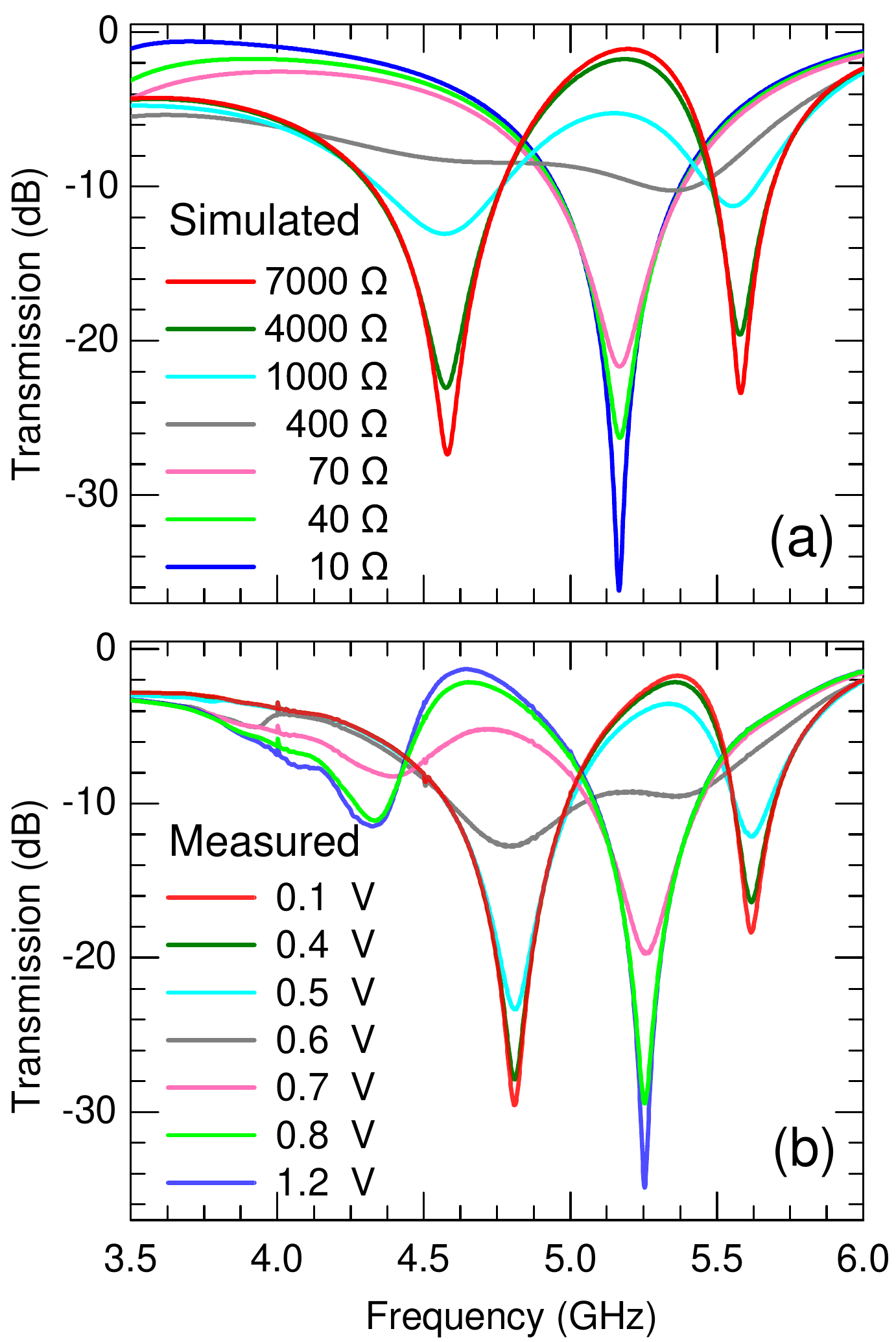}
\caption{Conductivity dependence of the transport through a metamaterial switch. (a) Plot of the simulated transmission spectra (in dB) for a serious of lumped resistance (from 7000 $\Omega$ to 10$\Omega$). (b) Measured transmission of the metamaterial switch sample under different biasing voltage changing from 0.1 V to 1.2 V.}
\end{figure}

To realize the inductive control of the EIT for switching, here we employ a PIN diode to implement the electrically tunable connection of the paired wires, a PIN-diode was inserted in between the pair of wires. As we know, the diode will be under a high resistance state when it is not biased or the voltage is small, because the PN junction blocks the flow of the carriers. However, when the bias voltage increases to some certain level, the drag ability on carriers of P-N junction decreases remarkably and the diode becomes of low resistance. The photograph of a sample fabricated for experimental investigation is presented in Fig. 1(b), two metallic conductors are welded on the magnetic wire pair for electric biasing the PIN diode. All the experiments were done in a standard C-band waveguide (model: WR-187), the scattering parameters measurement were carried out using a vector network analyzer(model: AV3629D). In the experiment, the diode was connected to an accurately tunable source to control the bias voltage.

Figure 3(b) presents the measured transmission of the metamaterial sample under different bias voltage changing from 0.1 V to 1.2 V. For not biased or low bias voltage situations, for example bias voltage of 0.1V, the transmission spectrum shows a peak around 5.25 GHz. Then we increased the bias voltage on the diode, it can be seen that the peak around 5.25 GHz changes dramatically to a transmission dip of about -35 dB during the bias voltage increasing from 0.1V to the 1.2V. The measured electric switch behavior at the EIT peak frequency agrees well with our theoretical prediction as presented in Fig. 3(a) (in which the values of resistance were chosen to compare with the measured spectra), indicating that the inductive modification on the resonance of magnetic wire pair can be used for practical modulating application. We also notice that the two dips associated with the EIT peak vanished gradually as the bias voltage increasing as the calculated results. However, an additional dip appears on the transmission spectrum around 4.4 GHz for higher bias voltages like 0.7 V, 0.8 V, and 1.2 V. We confirmed that the unexpected dip is due to the introduce of the biasing arms [red wires in Fig. 1(b)], the biasing arms extended the size of paired wires and thus shifted the magnetic resonance of connected wire pair to about 4.4 GHz (the paired wires are almost electrically connected under high biasing voltages, and thus the resistance of the diode is rather low), the biasing arms also makes the measured working frequencies are higher than the prediction without considering the biasing. To make a short summary, although the measured results is slightly different from the predicted results [see Fig. 3(a)] of theoretical model which is due to the biasing in measurements, our experiments generally validated the feasibility of our proposal that the novel EIT phenomenon can be controlled to realize metamaterial modulators in a subwavelength scale.

\begin{figure}[h]
\centering\includegraphics[width=3.5in]{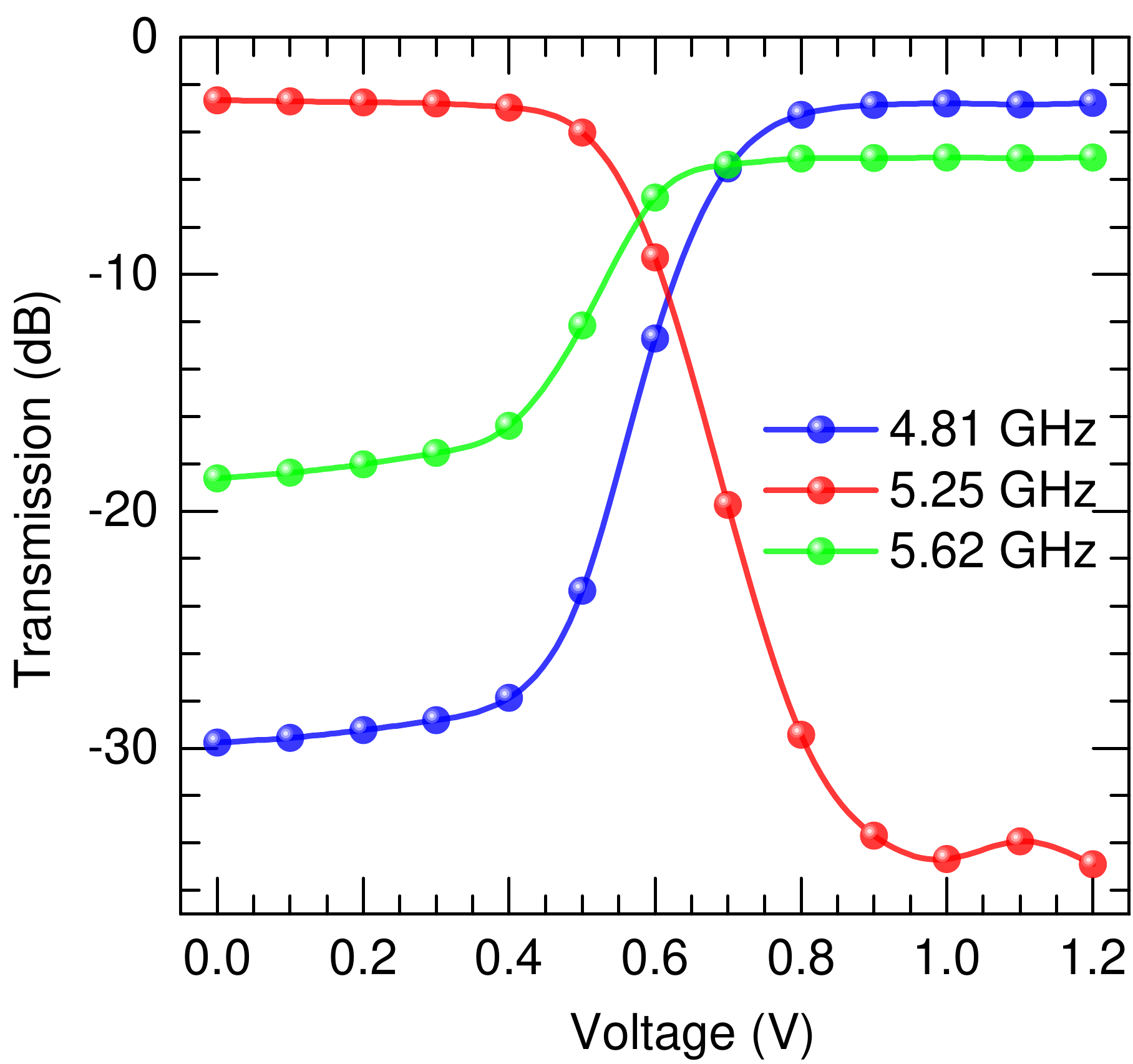}
\caption{Measured transmission (in dB) versus biasing voltage on the diode (from 0 to 1.2 V) at the three switching frequencies. The transmission are plotted for the EIT peak frequency 5.25 GHz (red), EIT dip frequencies 4.81 GHz (blue) and 5.62 GHz (green).}
\end{figure}

To show more details on the working performance of our metamaterial modulator, we plot in Fig. 4 the measured transmission as a function of the bias voltage at three characteristic frequencies of the EIT spectrum or frequencies at which remarkable modulation in transmission can be achieved through electric biasing, that transmission at 5.25 GHz, 4.81 GHz, and 5.62 GHz. The modulation curve at 5.25 GHz, shown as red curve in Fig. 4, shows a range with bias voltage from 0.4 V to 0.9 V, during which the transmission drops rapidly from the high transmission region to low transmission region with modulation ratio of about 31 dB. The modulations at 4.81 GHz and 5.62 GHz show changing in the same bias voltage range, but the transmission rises from low transmission level to high transmission level as the bias voltage increases, with modulation ratio of about 27 dB and 14 dB, respectively. Generally speaking, we find that the transmission can be greatly under a small change in bias voltage at multiple frequencies thanks to the novel EIT spectrum and the sensitive tunability of the magnetic resonance of a wire pair.

In summary, we experimentally demonstrated an electrically controlled metamaterial modulator by tuning the formation process of EIT like spectrum. Since that the EIT spectrum comes from the coherent interference of a sharp resonance with a background of broaden response, we proposed to control the EIT spectrum by introducing a PIN diode between a pair of wires, the resistance of the diode can be electrically tuned for inductively controlling the magnetic mode of the paired wires and thus the transmission spectrum. As a consequence, the transmission of electromagnetic waves in the three characteristic bands on the EIT spectrum can be effectively modulated by the voltages on the diode. The experiments confirmed that the transmission of the metamaterial switch can be significantly modulated under rather small biasing voltages at the characteristic peak/dips of EIT like spectrum. The finding of the multi-frequency modulating has benefits for translating novel spectral behaviors in metamaterials into practical applications in manipulating light.

The authors would like to acknowledge financial support from the National Science Foundation of China (NSFC) (Grant Nos. 61505164 and 11372248), the Program for Scientific Activities of Selected Returned Overseas Professionals in Shaanxi Province, the Fundamental Research Funds for the Central Universities (Grant Nos. 3102015ZY079 and 3102015ZY058), the Shaanxi Project for Young New Star in Science and Technology (Grant No. 2015KJXX-11),and National Undergraduate Innovative Program (Grant Nos. 201510699212 and 201510699213).

\end{document}